\documentclass[useamsmath,useAMS,usenatbib,usegraphicx,usedcolumn]{mn2e}

\title[Stellar populations in Fossil BCGs]{Spatially resolved stellar population parameters in the BCGs of two fossil groups}
\author[Proctor, Mendes de Oliveira, \& Eigenthaler]
{Robert N. Proctor$^{1,2}$, Claudia Mendes de Oliveira$^{1}$, Paul Eigenthaler$^{3}$\\
$^1$  Universidade de S\~{a}o Paulo, IAG, Rua do Mat\~{a}o, 1226, S\~{a}o Paulo, 05508-090, Brasil\\
$^2$  Observat\'{o}rio Nacional, Rua Gal. Jos\'{e} Cristino, 20921-400, Rio de Janeiro, Brazil\\
$^3$  Instituto de Astronom\'{i}a y Astrof\'{i}sica, Pontificia Universidad Cat\'{o}lica de Chile, Av.\ Vicu\~{n}a Mackenna 4860, Santiago, Chile\\
email: rnp059@gmail.com}
\begin{document}


\maketitle

\label{firstpage}

\begin{abstract}
We report the results of Gemini/GMOS long-slit spectroscopic
observations along the major and minor axes of the central galaxies in
two fossil groups, SDSS J073422.21+265133.9 and SDSS
J075828.11+374711.8 (the NGC 2484 group).  Spatially resolved
kinematics and stellar population parameters (ages, metallicities and
$\alpha$-element abundance ratios) derived using $\sim$20 Lick indices
are presented.  Despite remarkable similarities in their morphologies,
photometric properties (luminosity and colour) and kinematics, the two
galaxies exhibit significantly different stellar population
parameters.

SDSS J073422.21+265133.9 exhibits a strong metallicity gradient
($\Delta$ [Z/H]/$\Delta$ R $\sim$ -0.4) all the way into the centre of
the galaxy. It also exhibits an age profile that suggest a relatively
recent, centrally concentrated burst of star formation superimposed on
an older, more spatially extended population.  NGC 2484, a well known
X-ray AGN, exhibits a flat core-like structure in its metallicity
gradient, but no detectable age gradient. The $\alpha$-element
abundance ratio ([E/Fe]) profiles of the two galaxies are also
significantly different. SDSS J073422.21+265133.9 exhibits a slightly
positive gradient ($\Delta$ [E/H]/$\Delta$ R $\sim$ 0.1), perhaps
again suggesting a more recent central burst of star formation, while
NGC 2484 shows a negative gradient ($\Delta$ [E/H]/$\Delta$ R $\sim$
-0.1), indicating that star formation may have happened "inside out".

Our analysis of these two galaxies of similar mass, morphology and
kinematics therefore suggests two different mechanisms to have been in
action during their formation.  Consequently, we conclude that the
central galaxies of fossil groups can not be considered a homogeneous
group with regard to their formation processes or star formation
histories.
\end{abstract}


\section{Introduction}
Fossil galaxy groups are defined as X-ray luminous
($L_{X,\,\textrm{\scriptsize bol}} > 0.5 \times 10^{42}$ $h_{50}^{-2}$
erg~s$^{-1}$) galaxy aggregates with a greater than 2 magnitude gap
between the brightest and second brightest galaxies within half the
virial radius \citep{jones03}. The commonly stated paradigm for the
formation of these systems is that the large magnitude gap is
generated by dynamical friction acting on the massive $L^{*}$ galaxies
near the centre of the group, which causes them to spiral inwards and
merge with the central galaxy. For the magnitude gap to be sustained,
this scenario requires the group undergoes no significant recent
mergers or infall events. According to this paradigm, fossil groups
therefore represent highly evolved, but otherwise undisturbed,
examples of galaxy groups that formed early in the history of the
universe. Hence the study of fossil groups has become of great
interest, as this scenario suggests that they represent the
undisturbed end-product of galaxy group evolution. Contrary to this
hierarchical merging scenario it was also suggested that fossil groups
could simply be \emph {failed groups} that formed with a top-heavy
luminosity function absent of $L^{*}$ galaxies (Mulchaey \& Zabludoff
1999). As such they represent an important benchmark against which
other systems can be compared.

Although fossil groups as whole systems are believed to have collapsed
early and have assembled most of their virial masses at higher
redshifts, in comparison with non-fossil groups
\citep{donghia05,dariush07}, first-ranked galaxies in fossil groups
may have merged {\it later} than non-fossil bright central galaxies
\citep{diazgimenez08}. The question to which this work is addressed is
then: does the formation process of fossil groups create significant
differences in the stellar population gradients of their first-ranked
galaxies when compared to non-fossil bright cluster galaxies
(hereafter BCGs)?

There are relatively few studies on the stellar population gradients
of BCGs, and for fossil groups, only one study has been carried-out so
far, concluding that first-ranked galaxies in fossil groups show
comparatively flat metallicity gradients indicative of a major merger
origin \citep{eigenthaler13}. \citet{loubser} presented a systematic
study of the relations between the stellar population gradients of 24
BCGs in non-fossil clusters and properties of the cluster where they
reside (richness, mass, etc). This study may serve as a benchmark
against which results on the properties of first-ranked galaxies of
fossil groups can be compared. Other smaller samples which focused on
the detailed study of stellar populations of BCGs are
\citet{brough07,spolaor,gorgas90,fisher95,sanchezblazquez06,mehlert,carter,davidgegrinder95}.
Studies of BCGs in non-fossil groups have shown that these galaxies
possess central stellar populations that are generally old, of super
solar metallicity and moderately $\alpha$-element enhanced, i.e.  they
exhibit similar ages and metallicities as non-BCGs of the same mass.
However, some studies hint towards higher [E/Fe] values in BCGs
\citep{linden07,loubser09}.  BCGs exhibit fairly strong negative
metallicity gradients, and shallow age gradients, both positive and
negative. Where available $\alpha$-element abundance ratios are shown
to be relatively shallow \citep{loubser}.  However, it is not simple
to infer a consistent formation history for BCGs that can at the same
time explain the gradients in the three parameters, age, metallicity,
and abundances. In any case, what is evident is that the dispersions
in the stellar population parameters are high, indicating a number of
possible formation scenarios for BCGs that depend on the details of
the merging history for each galaxy.

Given our small sample, a full statistical comparison of fossil to
non-fossil BCGs is not possible in this paper, but must wait until a
larger sample of first-ranked fossil group galaxies have been studied
at the same level of detail. In this work we therefore simply continue
the accumulation of stellar population parameters age, [Fe/H], [E/Fe]
(a proxy for [$\alpha$/Fe] \citep{proctorsansom02}), and [Z/H] for
BCGs in fossil groups.

The paper is organised as follows. In Section \ref{observations} the
sample selection, observations and data reduction are described.
Section \ref{kinematicsstellarpopulations} presents galaxy kinematics
and stellar population parameters.  Section \ref{discussion} discusses
the obtained results while Sect.\ \ref{conclusions} shows our
conclusions.

\section{Sample  selection,  observations  and  data  reduction} 
\label{observations}

We report on Gemini GMOS spectroscopic data for the BCGs of the two
fossil groups SDSS J073422.21+265144.9 and SDSS J075828.10+374711.8
(hereafter SDSS0734 and SDSS0758).  The sample was selected from the
list of fossil groups in \citet{diazgimenez08}, originally identified
in the Sloan Digital Sky Survey (SDSS) DR6 \citep{adelman08}.
\citet{diazgimenez08} used a Friends-of-Friends analysis to find
groups within the SDSS with an over-density contrast of 200, more than
10 spectroscopically confirmed members, masses greater than
5$\times$10$^{13}$ $h^{−1}$ M$_{\odot}$ and redshifts lower than 0.1.
Using these criteria, \citet{diazgimenez08} identified five groups
with $\Delta$m$_{12}$ values larger than 2.0. We had originally
intended to observe all five bona-fide fossil groups from
\citet{diazgimenez08}, however, given the time allocation constraints,
we obtained data for only two of them (galaxies II and IV in
\citealt{diazgimenez08}).  The two galaxies possess nearly identical
photometric properties and reside in groups with very similar
properties. Their $g$ band magnitudes, $g-r$ colours, sizes,
minor/major axis ratios ($b/a$), and redshifts are given in Table
\ref{photometricdata}. The $\Delta$m$_{12}$ values, velocity
dispersions, number of spectroscopically confirmed members within the
virial radius and richness of the groups in which these BCGs reside
are also given in Table \ref{photometricdata}. Magnitudes, colours and
redshifts were taken from the SDSS DR9.  The absolute $g$ band
magnitude was taken from the SDSS DR9 {\small photoz} table. For
effective radii, we considered the isophotal radius at $\mu_{K}$=20
mag arcsec$^{-2}$ from the 2MASS extended objects catalogue
\citep{2MASS} to be a good estimate (see also
\citealt{proctor08}). Based on the photometric similarities of both
galaxies and the groups in which they resdie, the selected targets are
ideal for studying the uniformity of the formation scenario for fossil
galaxy groups.  Hence we carried out spectroscopic observations along
the major and minor axes of these two fossil BCGs to study their
spatially resolved stellar population parameters.

\begin{table}
\caption{Overall properties of the observed fossil BCGs and the groups in which they reside.}
\begin{center}
\begin{tabular}{ l  . . }
\hline
BCGs                       &            \multicolumn{1}{c}{SDSS0734}            &             \multicolumn{1}{c}{SDSS0758}             \\
\hline                                                                                                                                    
$\alpha_{\rm J2000}$        &  \multicolumn{1}{c}{$07^{\rm h}34^{\rm m}22\fs2$}  &   \multicolumn{1}{c}{$07^{\rm h}58^{\rm m}28\fs1$}   \\
$\delta_{\rm J2000}$        &  \multicolumn{1}{c}{$+26\degr51\arcmin44\farcs9$}  &   \multicolumn{1}{c}{$+37\degr47\arcmin11\farcs8$}   \\
$g$ [mag]                   &                        15.35                       &                         13.87                        \\
$M_g$ [mag]                 &                       -22.67                       &                        -22.58                        \\
$g-r$ [mag]                 &                         1.00                       &                          0.89                        \\
$r_{\rm eff}$ [arcsec]      &                        15.2                        &                         30.2                         \\
$r_{\rm eff}$ [kpc]         &                        24.1                        &                         25.7                         \\
$b/a$                       &                         0.80                       &                          0.78                        \\
$z$                         &                         0.0796                     &                          0.0408                      \\
\hline
Groups                        &            \multicolumn{1}{c}{SDSS0734}            &             \multicolumn{1}{c}{SDSS0758}             \\
\hline
$\Delta$m$_{12}$ [mag]                   &   2.01                     &    2.64        \\
$\sigma$ [km s$^{-1}$]                     &   551                     &     563       \\
Members                     &   16                     &      38      \\ 
Richness                    &    5                    &        4    \\
\hline

\end{tabular}
\end{center}
{\bf Notes:} Magnitudes, colours and redshifts were taken from the
SDSS DR9, while coordinates, minor/major axis ratios, and effective
radii were taken from NED. Members is the number of galaxies within the
virial radius with SDSS spectroscopy. Group richness is defined as
the number of spectroscopically confirmed members brighter than
0.4L$^*$.
\label{photometricdata}
\end{table}

The observations were  performed using GMOS  on Gemini North  on  December  24 in 2011 (GN-2011B-Q-107) in long-slit  mode.  Observations  were
carried out  using the   B600+G5307 grating  and a   long-slit of  1 arcsec   slit-width, yielding  a dispersion  of 0.92  \AA~pixel$^{-1}$. We measured  a
 spectral resolution  of  4.7~\AA\ over  a  wavelength  range  of  3500 to  6300~\AA \space from  the  FWHM   of  multiple  arc calibration lines.  Three
exposures of 1800~s were performed along each  of the major and  minor axes  of both  galaxies. Exposures were dithered in the spectral direction in order 
to provide coverage across the GMOS chip gaps$^{1}$.   Arc  and flat-field observations were interleaved between  the target  exposures. The   bias  frames 
provided by   the observatory were   taken  from a  preceding run  on December 12 in 2011. The seeing was $\geq$1.2 arcsec.

\begin{figure*}
\centering
\includegraphics[width=17cm]{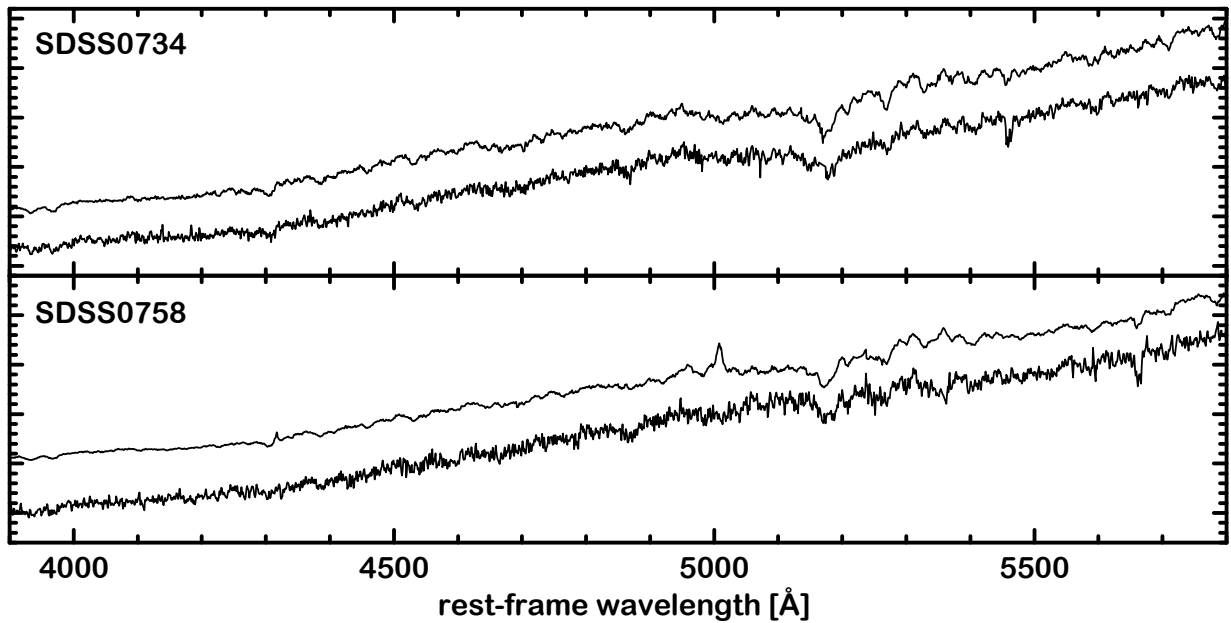}
\caption{\label{spectra} De-redshifted spectra along the major axis of
  the observed BCGs of the two fossil groups. The central bin and one in the
  outskirts around $\log(r/r_{\rm eff})\sim-0.3$ are shown.  Sky-line
  residuals have been removed for better visualisation. The spectra
  have been normalised to the flux at $\lambda$5000\AA. SDSS0758 shows
  prominent [OIII] $\lambda\lambda4959,5007$\AA \space emission in the
  center and is a well-known X-ray AGN \citep{sun09}.}
\end{figure*}

The data reduction was carried out using the {\small IRAF} Gemini
{\small GMOS} package and {\small STARLINK} software. Early in the
data reduction process, it became apparent that the data suffered from
significant contamination by scattered light, which had serious
repercussions for flat-fielding and sky-subtraction. In order to
minimise scattered light effects, the reduction procedure carried out
differed slightly from the usual sequence.  First, both target and
flat-field frames were bias subtracted, cleaned of cosmic rays and bad
pixels, and subsequently wavelength calibrated using Cu-Ar
comparison-lamp exposures.  Next we used the only regions in our
observations in which the scattered light could be measured directly,
i.e.\ the {\it bridges} in the long-slit\footnote{see the definition
of GMOS chip gaps and bridges on the Gemini homepage {\tt
http://www.gemini.edu/node/10663}}, to estimate scattered light
levels, again in both target and flat-field frames. We found that some
20\% of the light reaching the detector was in the form of scattered
light.  Estimates of the scattered light contamination were made by
interpolating between the two bridges, along the spatial
direction. These levels were then subtracted from each
exposure. Flat-fielding of the decontaminated target frames was then
carried out using the corrected flat-field frames.  The resultant
light profiles { along the spatial direction were found to be much
  flatter at large radii, i.e.\ more consistent with the constant sky
  level one would expect at such radii.  Sky subtraction was then
  performed in the usual manner, by fitting a linear function to
  background windows along the spatial direction on each side of the
  galaxy spectrum outside the effective radii. The data were then
  binned along each axis so that the outer bins still yielded a
  reasonable S/N. In the center, typically 2 pixels were binned while
  in the outskirts, around $r_{\rm eff}$, about 50 pixel rows were
  added.  We measure a typical S/N of $\sim 70$ in the center and
  $\sim 15$ in the outskirts. Figure \ref{spectra} shows spectra along
  the major axes for both galaxies. The central bin and one in the
  outskirts around $\log(r/r_{\rm eff})\sim-0.3$ are presented.

\section{Kinematics and Stellar population parameters}
\label{kinematicsstellarpopulations}

The recession velocities and velocity dispersions of each bin along
the major and minor axes were measured using the {\small IRAF fxcor}
routine.  \citet{bruzualcharlot03} SSP models that had been broadened
to the Gemini resolution of 4.7\AA\ were used as templates for this
procedure. The SSP model showing the highest cross-correlation peak
was selected for the final measures. However, variation in the values
derived using different templates was found to be less than the errors
quoted in our results. The resulting values are shown in
Fig.\ref{kins}. Radial velocities are shown with respect to the
systemic velocities. We find no evidence for rotation along either
axis of either galaxy.  The velocity dispersion profiles are both
shallow.  Indeed, there is no evidence for any reduction in velocity
dispersion with radius in SDSS0758. It is therefore clear that both
galaxies reside in substantial dark matter halos. The systemic
velocities of both galaxies and the measured central values of the
velocity dispersion are listed in Table \ref{slopes}.  Radial
velocities and velocity dispersions at each radial bin are given in
Appendix \ref{kinematicappendix}.

\begin{figure*}
\centering
\includegraphics[width=12cm]{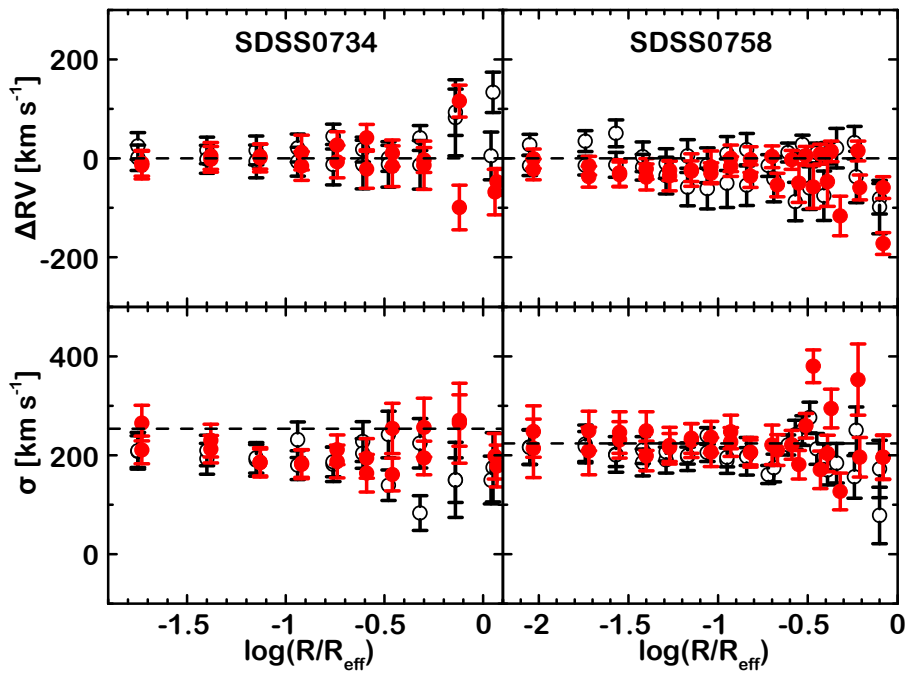}
\caption{\label{kins}Kinematics of the observed BCGs of the two fossil
  groups. Radial velocity and velocity dispersion profiles are
  shown. Red points are measured along the major axis while open ones
  present the minor axis.  The horizontal lines show the systemic
  velocities and central velocity dispersions. There is no evidence
  for rotation along either axis of either galaxy.  The velocity
  dispersion profiles are both shallow.}
\end{figure*}

\subsection{Measurement of stellar population parameters}
\label{sspmeasurement}
To derive stellar population parameters, Lick/IDS line-strength
indices as defined in \citet{worthey94} and \citet{worthey97} have
been measured. The data were corrected for internal velocity
dispersion as described in \citet{proctorsansom02}.  In the absence of
Lick standard star observations, full calibration to the Lick system
could not be performed. However, as noted in many previous studies
(e.g. \citealt{proctorsansom02}), the only indices that significantly
depend on such a calibration are Mg$_1$ and Mg$_2$. These indices were
therefore omitted from all further analysis. NaD was also excluded
because of its contamination by interstellar absorption. The observed
wavelength range did not cover the redshifted TiO indices. As a result
a total of 22 Lick indices from H$\delta$ to Fe5782 were measured with
20 of these used in our analysis. The measurement of stellar
population parameters age, [Z/H] and [E/Fe] was then carried out using
the multi-index $\chi^{2}$-fitting technique detailed in
\citet{proctorsansom02}. [Fe/H] was then derived from [Z/H] and [E/Fe]
using [Fe/H]=[Z/H]--0.94[E/Fe].  The 20 indices were fit, first
subject to three-sigma clipping. This removed indices contaminated by
effects such as sky-line residuals and chip gap residuals as well as
highlighting positions in the galaxies affected by emission lines (in
this case the central two pixels of SDSS0758 only). The stability of
the fits with respect to the exclusion of the remaining indices was
then tested to ensure that no individual index was creating tension in
the fit.  In most cases the indices causing such tensions were found
to be remaining sky-line residuals, or other clear distortions in the
spectrum. Those indices were also excluded from the final fits. The
resultant fits generally included between 17 and 20 indices.  The
derived stellar population parameters are shown in Fig. \ref{agez}.

\begin{figure*}
\includegraphics[width=12cm]{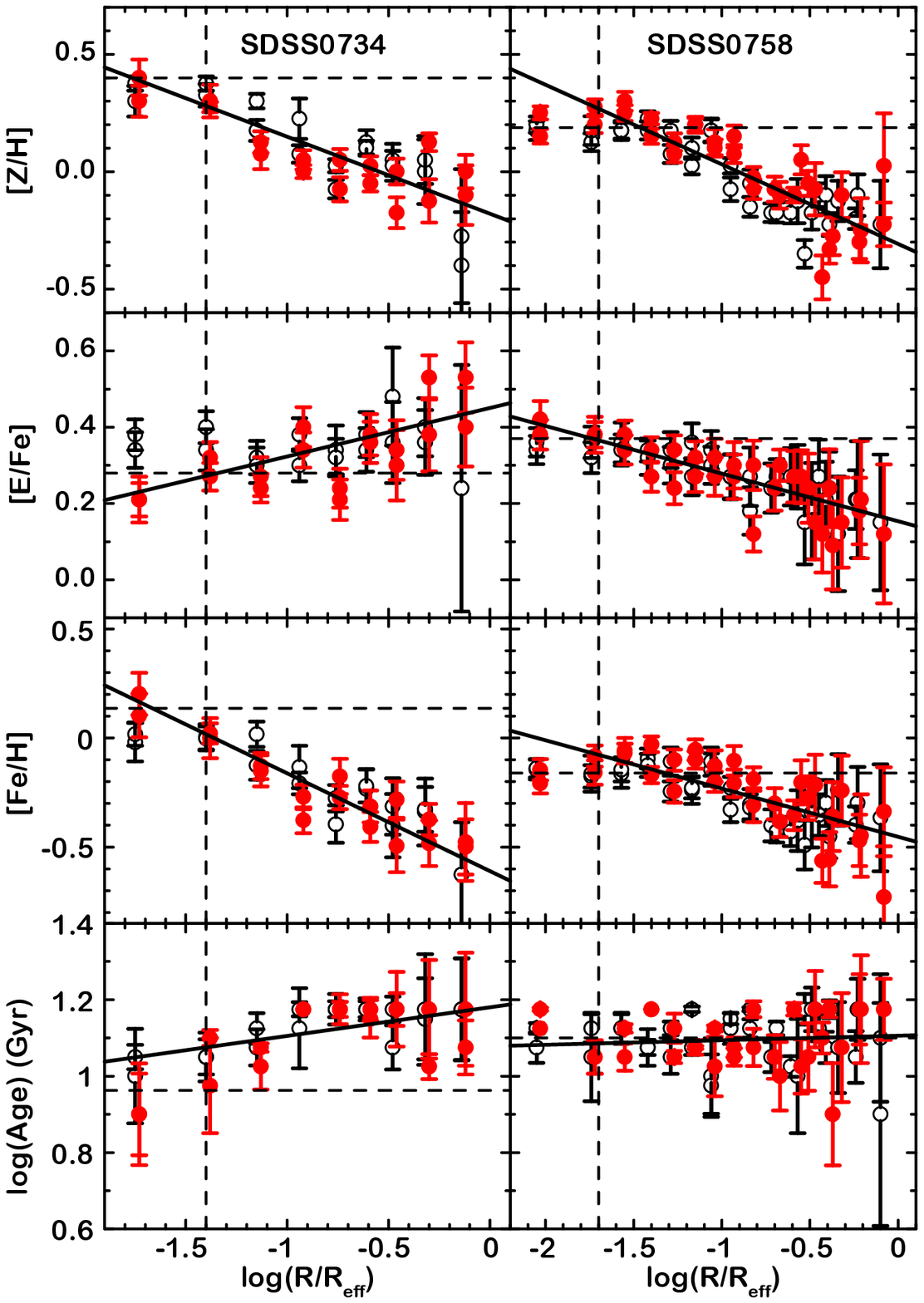}
\caption{Spatially resolved stellar population parameters for the
  observed BCGs of fossil groups. The data are measured with the
  multi-index $\chi^{2}$-fitting technique detailed in
  \citet{proctorsansom02}. Filled, red points are measured along the
  major axis while open, black points present the minor axis.  The
  horizontal dashed lines show the measured central values. The
  vertical lines indicate the typical seeing during our
  observations. Solid lines show linear least squares fits to the
  data.}
\label{agez}
\end{figure*}

\subsection{Comparison with full-spectrum fitting}
\label{fullspectrumfitting}
To verify the stability of our measurements, we also compared our
results from the multi-index $\chi^{2}$-fitting technique with a
full-spectrum fitting technique.  Therefore, the open-source package
ULySS \citep{koleva09} was used to fit SSP models directly to the
observed galaxy spectra.  We utilized Pegase$\,$HR models resolved in
[$\alpha$/Fe] \citep{alphamodels} based on the Elodie$\,$3.2 library
of stellar spectra \citep{elodie32}\footnote{\tt
  http://ulyss.univ-lyon1.fr/optional.html}.  The library involves an
interpolator to provide a stellar spectrum at any point in the
parameter space ($T_{\rm eff}$, log~$g$, [Fe/H], and [$\alpha$/Fe]).
All Pegase$\,$HR models are computed assuming a Salpeter IMF and
Padova 1994 evolutionary tracks providing synthetic SSPs with ages
between 1$ -$20000$\,$Myr, metallicities between $-2.3$ and
$0.69\,$dex, and [$\alpha$/Fe] enhancement between 0 and 0.4.  A
multiplicative polynomial is used to adjust the overall spectral shape
to the SSP model. To achieve reasonable values an order of 40 was
chosen.  The ULySS parameter {\tt \textbackslash CLEAN} was considered
in the fitting procedure to exclude possible outliers in the galaxy
spectrum, resulting from any remaining night sky emission.  200
Monte-Carlo simulations were computed for every spectral bin,
repeating a fit successively with random Gaussian noise. The dimension
of the added noise was based on a user-defined signal-to-noise ratio
provided for each spectral bin.  The S/N ratios were measured with
{\small IRAF splot} at a rest-frame wavelength of $\sim5140\,$\AA. The
resulting point distributions were then used to calculate average SSP
ages, [Fe/H] (see also \citealt{eigenthaler13}) and
[$\alpha$/Fe] values. [Z/H] was then also calculated from these data 
using [Z/H]=[Fe/H]+0.94[E/Fe]. Outliers clearly detached from the main point
distributions have been excluded for these measurements. Error bars
were estimated as the standard deviation of the point
distributions. For the comparison, we analysed the major axis spectra
of both galaxies with ULySS, and averaged the stellar population
parameters from both galaxy sides. 

Besides stellar population parameters, ULySS also allows us to measure
the kinematic properties of our galaxies since the package measures
the shift in wavelength and broadening of each SSP model to match the
observed galaxy spectrum. ULySS determines the broadening relative to
the SSP model dispersion $\sigma_{\rm{model}}$, amounting to 13 km
s$^{-1}$ for the Pegase HR models. Taking into account instrumental
dispersion, velocity dispersions are computed via the relation: $\sigma                                                                                      
_{\rm{phys}}^2 = \sigma _{\rm{ulyss}}^2 + \sigma _{\rm{model}}^2 -                                                                                          
\sigma _{\rm{instr}}^2$.

We also investigated the use of composite stellar populations within ULySS
by attempting to fit the {\it relatively} high S/N central bins of each
galaxy with a combination of two independent SSPs. However, even in
these central bins the S/N was not sufficient to permit the
convergence of the fits to a stable solution given the large number of
free parameters in such a fit. Therefore we report here only fits
utilising single SSP models.

Figure \ref{comparison} shows the comparison of our results from the
multi-index $\chi^{2}$-fitting technique with the corresponding values
derived from ULySS full-spectrum fitting.  We confirm the kinematic
properties of both galaxies. However, we note that our ULySS results
show systematically larger values for velocity dispersion with an
average offset of $\sim 30$ km s$^{- 1}$ compared to the values
derived from the {\small IRAF fxcor} routine. The cause of this slight
discrepancy remains unidentified.  However, the offset (which is only
slightly larger than the velocity dispersion errors) was found to have
no significant effect on the stellar population parameters derived.

Qualitatively, both the kinematic and stellar population trends show
fair agreement. However, there are some noticeably quantitative
differences in the results of the stellar population analyses. Most
noticeably, the [E/Fe], although
showing similar gradients, are systematically lower in the ULySS full
spectral fitting results. There are also some slight systematic
differences in the derived [Fe/H] values. For SDSS0734 we confirm the
negative [Fe/H] gradient but find a shallower slope, while for
SDSS0758, we confirm the flat inner metallicity gradient, but find a
flat gradient also at larger galactic radii. Interestingly, however,
the derived [Z/H] values (which are a combination of [Fe/H] and
[E/Fe]) show good agreement at all points except for the two high
[Z/H] inner points of SDSS0734. It is therefore evident that these
differences arise as the result of the differing model set used
(Thomas, Maraston \& Bender in the multi-index fitting analysis and
Pegase$\,$HR models in the ULySS full spectral fitting
analysis). With regards age, for SDSS0734 agreement is very
good. However, the age gradient for SDSS0758 shows the largest
discrepancy in our comparison. While we find a flat age gradient from
multi-index fitting, we obtain systematically lower ages from
full-spectrum fitting which even show a negative trend around
$\log(r/r_{\rm eff})\sim-0.7$. We note that in this galaxy the
deviations in age and [Fe/H] follow the age-metallicity
degeneracy, i.e. higher [Fe/H] is compensated for by younger age.
It is our intention to continue to make these comparisons in future
data analyses in the hope of identifying consistent systematic
difference between the model sets.

\begin{figure}
\centering
\includegraphics[width=\columnwidth]{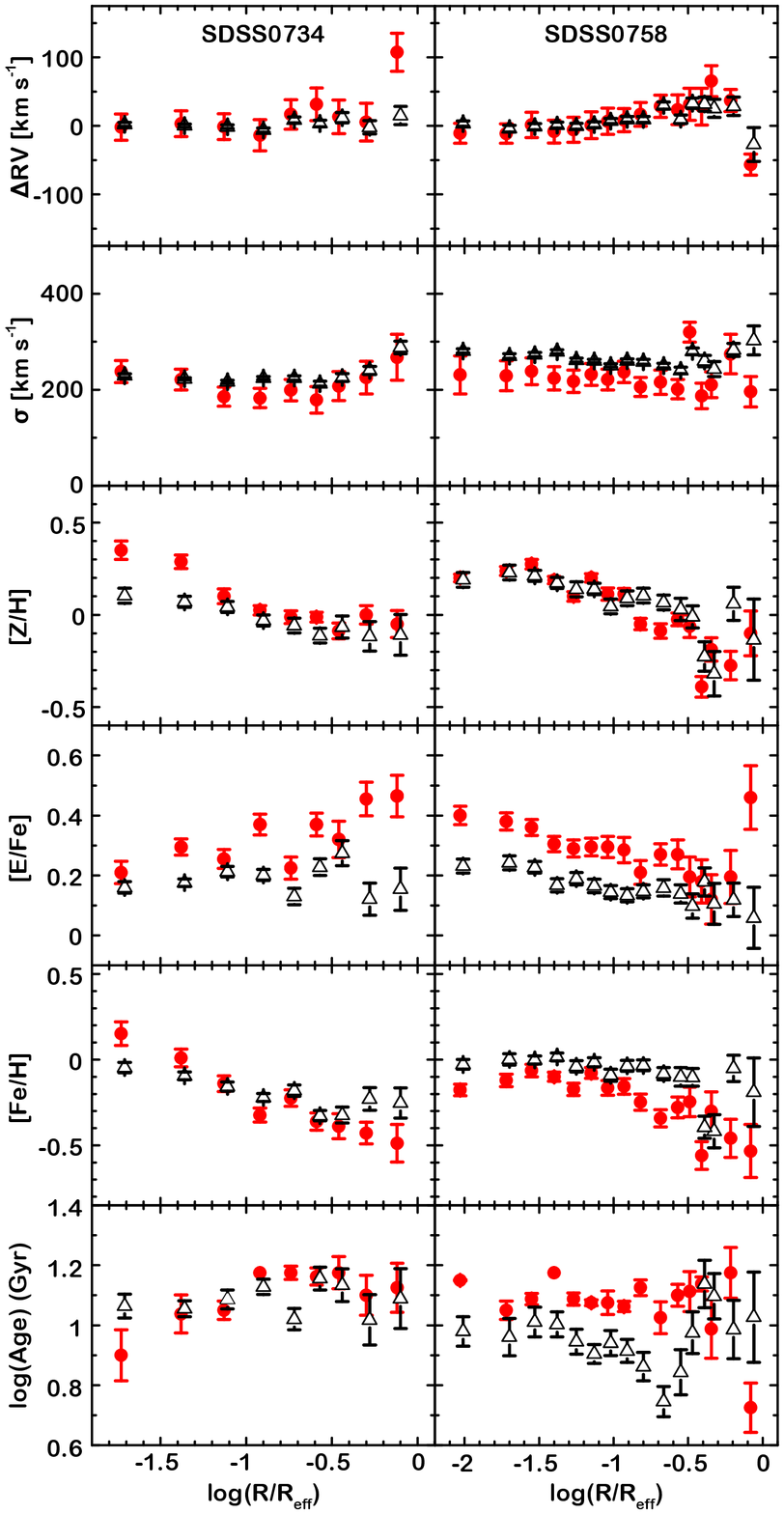}
\caption{\label{comparison}Comparison between the results from the
  multi-index $\chi^{2}$-fitting technique detailed in
  \citet{proctorsansom02} and the full-spectrum fitting technique
  ULySS. Red symbols show the major axis data from Fig.\ref{agez},
  while open triangles show the results from full-spectrum fitting. In
  both cases the data have been folded along the major axis.}
\end{figure}

\section{Discussion}
\label{discussion}
Figs. \ref{agez} and \ref{comparison} show the spatially resolved
stellar population parameters resulting from our analysis.  It can be
seen from Figure \ref{agez} that there is good agreement between
values on either side of each axis and between the major and minor
axes of each galaxy. This evidences the robustness of the multi-index
fitting technique that we employ.  However, we note that there is a
hint of ellipticity in the [Z/H] and perhaps [Fe/H] profiles of
SDSS0758 such that they decline faster along the minor axis. SDSS0758
has a minor/major axis ratio of $\sim$0.8, which would cause a
displacement in log radius of points along the minor axis of -0.1 with
respect to the major axis values, consistent with the small
displacement evident in Fig.\ \ref{agez}. We detect no such
displacement in SDSS0734, despite it exhibiting a similar ellipticity
to SDSS0758.

The aim of this work is to begin the process of comparing the star formation
histories and/or merger histories of the BCGs of fossil systems to those of
normal systems. As well as the comparison of the central ages and metallicities of
such systems, another, and potentially more powerful, approach is the comparison
of the gradients in these parameters. As a result there has been much recent
interest in the use of metallicity gradients as a probe of the merger histories
of galaxies. However, such comparisons are severely hampered by the number of free
parameters that describe the star formation and merger histories of large
early-type galaxies, such as the BCGs in our sample. Key amongst these
parameters are whether the mergers that the central galaxies experiences are
¨wet¨ or ¨dry¨ (whether or not there is significant gas present during the
merger).

In the case of dry mergers the metallicity gradient of the merger remnant
depends on both the mass ratio of the merger event and on the pre-merger
gradients of the galaxies involved (e.g. \citealt{dimatteo09}). Conceivably,
age gradients could also be produced in the case of dry mergers if the two
merging galaxies have differing pre-merger ages.

For wet mergers, the effects on age gradients are generally more clearly
defined, since the gas present in the merging galaxies is known to be funneled
towards the galactic centers during the merger (e.g. \citealt{rupke10}), where
bursts of star formation then take place (e.g. \citealt{ellison13}). These
central starbursts then induce positive age gradients (younger towards the
centre). However, the prediction of metallicity gradients in merger remnants is
\emph{much} more complicated, since the gas driven to the centre of the merger
remnant can drive the gradient either up or down depending on the specifics of
the quantity, source and metallicity of the gas \citep{torrey12}.

Clearly, the interpretation of the gradients in individual galaxies
is, at best, a complex process.  This underlies the stated aim of this
work - to begin the process of measuring gradients in fossil group
BCGs in order to ultimately perform a \emph{statistical} comparison to
the BCGs of normal systems.  In the following we therefore refrain
from attempting detailed interpretation of the individual galaxies,
confining ourselves to general comments and comparisons between the
individual galaxies.\\

Gradients of the stellar population parameters were obtained by
fitting the radial profiles with linear relations of the form;

\begin{displaymath}
X[r/{r_{{\rm{eff}}}}] = {X_{{\rm{eff}}}} + {\nabla _{X}}\log\left(  {r/{r_{{\rm{eff}}}}} \right),
\end{displaymath}

\noindent where $X$ denotes the corresponding parameter.  The fits
have been weighted by the errors associated with each value.
Gradients were computed excluding the seeing affected innermost
point. We note that our derived gradients in general fit the data well
out to the effective radius.  The computed gradient slopes $\nabla
_{X}\equiv\Delta X/\Delta\log(r/r_{\rm eff})$ are given in Table
\ref{slopes}. The significance of these gradients were tested using
their correlation coefficients. All were found to be of greater than
3$\sigma$ except for the [E/Fe] in SDSS0758 (only 2$\sigma$) and log(age)
in SDSS0758 which in any case we find to exhibit no discernible
gradient.  The gradients in both galaxies were also found not to be
significantly affected by the exclusion of the low signal-to-noise
outer points.

We can readily see that despite the photometric
similarities between the two galaxies, there are significant
differences in the radial gradients of their stellar population
parameters.

Fig. \ref{agez} shows that, in SDSS0734 there is clear evidence that
the central region of the galaxy is younger than its outer
regions. There is also a suggestion that the central region has lower
[E/Fe] than the outer regions. Both [Fe/H] and [Z/H] show negative
gradients over the whole radial range probed.  This galaxy therefore
appears to have formed either in a single burst, but star formation
with the central regions continuing for 1-2 Gyr longer than in the out
regions i.e. long enough for Type Ia supernovae to begin driving the
central [E/Fe] down, or the galaxy may have formed an old population
upon which a central burst was then superimposed a few Gyr later. In
either case we might characterise the central regions of this galaxy
as having formed over a relatively extended period of time.

In SDSS0758, on the other hand, the galaxy exhibits no detectable age
gradient but a negative [E/Fe] gradient.  Although the galaxy also
exhibits negative [Fe/H] and [Z/H] gradients, there is a strong
suggestion of a flattening in this gradient in the central
regions. Indeed the [Fe/H] gradient appears to be flat all the way out
to $\log(r/r_{\rm eff})=-1$, and from our full-spectrum fitting
results, we measure a flat gradient even out to $\log(r/r_{\rm
  eff})=-0.5$.  The [E/Fe] gradient in this galaxy suggests that this
galaxy also formed over a relatively extended period of time, except
in this case the later star formation took place in the outer regions,
rather than in the central regions as was the case in SDSS0734. This
galaxy therefore resembles a bulge and disc galaxy in which star
formation was truncated very early, but slightly later (perhaps
$\sim$1 Gyr) in the outer regions. This is supported by the flat
(perhaps even negative) [Fe/H] gradient in the central region which
can arise if star formation in the central was truncated prior to the
star formation further out.

It is the aim of this project to compare the BCGs of fossil galaxy
groups with BCGs in non-fossil aggregates.  However, with a sample of
only two galaxies we can draw no firm conclusions at this point, other
than to note that these galaxies possess stellar population parameters
and gradients consistent with the range of values exhibited by normal
BCGs \citep{loubser}.

\begin{table*}
\begin{centering}
\caption{Central values and radial  gradients of the measured  stellar population parameters. System  velocities and central velocity  dispersions are
also given. Errors are listed below the corresponding values.}
\begin{tabular}{c..c.ccc...}
\hline
            &   \multicolumn{1}{c}{RV}    &  \multicolumn{1}{c}{$\sigma_{0}$}  &  \multicolumn{1}{c}{log(age)$_{0}$}  &  \multicolumn{1}{c}{[Fe/H]$_{0}$}   &  \multicolumn{1}{c}{[E/Fe]$_{0}$}  &   \multicolumn{1}{c}{[Z/H]$_{0}$}   &            $\nabla$log(age)             &   \multicolumn{1}{c}{$\nabla$[Fe/H]}  &  \multicolumn{1}{c}{$\nabla$[E/Fe]}  &  \multicolumn{1}{c}{$\nabla$[Z/H]}   \\
\hline                                                                                                                                                                                                                                                                                                                                                                                                   
  SDSS0734  &            23868            &                254                 &                 0.96                 &                 0.14                &                0.28                &                0.40                 &                  0.08                   &                 -0.45                 &                 0.13                 &                -0.33                 \\
            &                6            &                 17                 &                 0.01                 &                 0.01                &                0.04                &                0.03                 &                  0.05                   &                  0.07                 &                 0.07                 &                 0.10                 \\
                                                                                                                                                                                                                                                                                                                                                                                                         
  SDSS0758  &            12209            &                224                 &                 1.10                 &                -0.16                &                0.37                &                0.19                 &                  0.01                   &                 -0.22                 &                -0.13                 &                -0.34                 \\
            &                9            &                  9                 &                 0.03                 &                 0.02                &                0.03                &                0.01                 &                  0.06                   &                  0.09                 &                 0.05                 &                 0.09                 \\
\hline
\end{tabular}
\label{slopes}
\end{centering}
\end{table*}

\section{Conclusions}
\label{conclusions}

We have presented spatially resolved stellar population parameters
log(age), [Fe/H], [Z/H] and [E/Fe] for two first-ranked fossil group
galaxies, SDSS0734 and SDSS0758.  These two galaxies have quite
different stellar population parameters, despite their similarities in
morphology, absolute magnitude, colour and central velocity
dispersions, as well as the groups in which they resdie. We suggest
that one of them, SDSS0734, may have had a recent central burst of
star formation superimposed onto an old stellar population, which
explains a positive age gradient, a steep negative metallicity
gradient and positive [E/Fe] gradient. For the other galaxy, SDSS0758,
however, a well-known X-ray AGN \citep{sun09}, we argue that the star
formation was complete $\sim$ 10 Gyr ago. However, the negative
gradient in [E/Fe] and flat central metallicity profile suggest that
star formation may have continued slightly longer in the outer regions
than in the central regions of the galaxy (i.e.\ inside out
formation). Our results therefore suggest that there may be
considerable dispersion in the range of measured stellar population
parameters of first-ranked galaxies in fossil groups. We plan to study
a larger sample of these galaxies in detail in order to have a better
notion of how the central galaxies in fossil groups formed and to
investigate if there are any differences in their properties with
respect to the properties of non-fossil BCGs.

\section*{Acknowledgments}

This work is based on observations obtained at the Gemini Observatory,
which is operated by the Association of Universities for Research in
Astronomy, Inc., under a cooperative agreement with the NSF on behalf
of the Gemini partnership: the National Science Foundation (United
States), the Science and Technology Facilities Council (United
Kingdom), the National Research Council (Canada), CONICYT (Chile), the
Australian Research Council (Australia), Minist\'{e}rio da Ci\^{e}ncia
e Tecnologia (Brazil) and Ministerio de Ciencia, Tecnolog\'{i}a e
Innovaci\'{o}n Productiva (Argentina). This research made use of
NASA's Astrophysics Data System, as well as IRAF and STARLINK
software.  IRAF is distributed by the National Optical Astronomy
Observatories, which is operated by the Association of Universities
for Research in Astronomy, Inc.  (AURA) under cooperative agreement
with the National Science Foundation.  RNP acknowledges financial
support from the Brazilian agency FAPESP (program number
2008/57331-0). CMdO also acknowledges FAPESP (program number
2006/56213-9) and CNPq. PE acknowledges support from FONDECYT through
grant 3130485. We thank Mina Koleva for useful discussions on the
full-spectrum fitting package ULySS.

\appendix
\section{Kinematic properties}
\label{kinematicappendix}

\begin{table*}
\begin{centering}
\caption{Kinematic data for SDSS0734. Results are shown for both major and minor axes.}
\begin{tabular}{..c..c}
\hline                                                                                                                                                                                          
                                              &                                   \multicolumn{2}{c}{major axis}   &      &         \multicolumn{2}{c}{minor axis}                              \\
\hline                                                                                                                                                                                            
   \multicolumn{1}{c}{$\log(r/r_{\rm eff})$}  &   \Delta {\rm RV}       &       \multicolumn{1}{c}{$\sigma$}       &      &      \Delta {\rm RV}        &              $\sigma$                 \\
\hline                                                                                                                                                                                            
 0.07                                         &         -47(27)             &      178(42)                                 &      &            134(41)              &             175(70)                     \\
-0.12                                         &         116(32)             &      270(52)                                 &      &             82(77)              &             150(76)                     \\
-0.30                                         &         -3(25)             &       194(34)                                 &      &             41(25)              &              224(50)                      \\
-0.46                                         &         10(27)             &       161(33)                                 &      &              0(26)              &              242(47)                      \\
-0.59                                         &         41(27)             &       194(40)                                 &      &             18(25)              &              228(40)                      \\
-0.74                                         &         27(27)             &       187(33)                                 &      &             44(25)              &              182(35)                      \\
-0.92                                         &         -15(29)             &      182(29)                                 &      &             21(28)              &              232(36)                      \\
-1.13                                         &          1(28)             &       186(30)                                 &      &             16(29)              &              190(32)                      \\
-1.38                                         &          5(27)             &       230(33)                                 &      &             16(27)              &              193(31)                      \\
-1.73                                         &         -13(28)             &      265(36)                                 &      &             25(27)              &              208(32)                      \\
\multicolumn{1}{c}{{\rm Centre}}              &         12(25)             &       237(34)                                 &      &            -11(25)              &              270(32)                      \\
-1.73                                         &         -10(26)             &      211(28)                                 &      &              1(26)              &              209(37)                      \\
-1.38                                         &         -2(26)             &       212(28)                                 &      &             -2(28)              &              210(31)                      \\
-1.13                                         &          4(25)             &       185(27)                                 &      &             -6(33)              &              195(31)                      \\
-0.92                                         &         12(35)             &       184(28)                                 &      &             -7(29)              &              180(29)                      \\
-0.74                                         &         -7(33)             &       211(30)                                 &      &            -13(41)              &              187(30)                      \\
-0.59                                         &         -22(39)             &      164(38)                                 &      &            -12(49)              &              204(30)                      \\
-0.46                                         &         -16(41)             &      254(51)                                 &      &            -12(46)              &              139(31)                      \\
-0.30                                         &         -14(49)             &      256(59)                                 &      &            -13(49)              &               83(35)                      \\
-0.12                                         &         -99(45)             &      265(81)                                &      &              93(47)              &              150(45)                      \\
 0.06                                         &         -68(46)             &      198(46)                                 &      &              5(48)              &              150(48)                      \\
\hline                                                                                
\end{tabular}
\label{kinematic734}
\end{centering}
\end{table*}

\begin{table*}
\begin{centering}
\caption{Kinematic data for SDSS0758. Results are shown for both major and minor axes.}
\begin{tabular}{..c..c}
\hline                                                                                                                                                                                 
                                              &                   \multicolumn{2}{c}{major axis}                   &      &                     \multicolumn{2}{c}{minor axis}       \\
\hline                                                                                                                                                                                 
   \multicolumn{1}{c}{$\log(r/r_{\rm eff})$}  &   \Delta {\rm RV}       &       \multicolumn{1}{c}{$\sigma$}       &      &      \Delta {\rm RV}        &          $\sigma$          \\
\hline                                                                                                                                                                                 
-0.08                                         &          -172(22)           &                   196(44)                    &      &             -81(31)             &             78(57)             \\
-0.22                                         &            15(20)           &                   353(72)                    &      &              32(33)             &            156(43)             \\
-0.37                                         &            14(21)           &                   294(39)                    &      &              10(22)             &            170(30)             \\
-0.43                                         &             9(20)           &                   171(38)                    &      &              11(20)             &            213(31)             \\
-0.51                                         &             5(19)           &                   260(25)                    &      &              28(19)             &            250(30)             \\
-0.59                                         &            -3(19)           &                   221(29)                    &      &              12(19)             &            221(20)             \\
-0.70                                         &             3(22)           &                   221(40)                    &      &             -14(22)             &            161(18)             \\
-0.82                                         &            -2(26)           &                   206(25)                    &      &              18(33)             &            198(38)             \\
-0.93                                         &             0(27)           &                   227(29)                    &      &              10(34)             &            189(26)             \\
-1.04                                         &           -17(32)           &                   236(32)                    &      &             -14(35)             &            206(18)             \\
-1.15                                         &           -24(33)           &                   234(30)                    &      &               5(33)             &            197(28)             \\
-1.27                                         &           -36(30)           &                   216(28)                    &      &             -10(30)             &            190(26)             \\
-1.40                                         &           -38(26)           &                   198(30)                    &      &             -18(26)             &            211(27)             \\
-1.55                                         &           -30(27)           &                   231(37)                    &      &              51(27)             &            202(36)             \\
-1.72                                         &           -37(21)           &                   249(40)                    &      &              35(21)             &            223(38)             \\
-2.03                                         &           -22(21)           &                   248(52)                    &      &              28(21)             &            217(35)             \\
\multicolumn{1}{c}{{\rm Centre}}              &           -19(19)           &                   215(55)                    &      &              29(19)             &            232(33)             \\
-2.03                                         &            -1(20)           &                   214(59)                    &      &             -15(20)             &            216(35)             \\
-1.72                                         &           -15(19)           &                   209(48)                    &      &             -15(19)             &            218(30)             \\
-1.55                                         &           -33(25)           &                   246(42)                    &      &             -13(25)             &            200(23)             \\
-1.40                                         &           -21(21)           &                   249(39)                    &      &               3(30)             &            185(27)             \\
-1.27                                         &           -24(20)           &                   220(37)                    &      &             -35(37)             &            207(18)             \\
-1.15                                         &           -26(21)           &                   230(34)                    &      &             -58(38)             &            199(17)             \\
-1.04                                         &           -30(21)           &                   206(29)                    &      &             -61(41)             &            239(17)             \\
-0.93                                         &           -17(20)           &                   247(34)                    &      &             -50(49)             &            196(19)             \\
-0.82                                         &           -36(23)           &                   206(30)                    &      &             -54(41)             &            197(18)             \\
-0.67                                         &           -54(24)           &                   210(31)                    &      &             -42(46)             &            174(28)             \\
-0.55                                         &           -50(39)           &                   181(29)                    &      &             -87(39)             &            241(29)             \\
-0.47                                         &           -58(43)           &                   380(33)                    &      &             -60(43)             &            276(31)             \\
-0.39                                         &           -47(50)           &                   204(37)                    &      &             -75(50)             &            197(26)             \\
-0.32                                         &          -117(40)           &                   127(37)                    &      &              21(40)             &            184(40)             \\
-0.21                                         &           -59(25)           &                   196(40)                    &      &             -37(52)             &            251(47)             \\
-0.08                                         &           -59(22)           &                   196(45)                    &      &             -99(54)             &            172(58)             \\
\hline                                                       
\end{tabular}                                                
\label{kinematic758}                                         
\end{centering}                                              
\end{table*}

\section{Stellar population parameters}
\label{sspappendix}

\begin{table*}
\begin{centering}
\caption{Measured stellar population parameters for SDSS0734. Results are shown for both major and minor axes. Errors are given in brackets. Log(age) is expressed in Gyr.}
\begin{tabular}{cccccccccc}
\hline                                                                                                                                                                                                                                                                                                                                  
                                    &                                                   \multicolumn{4}{c}{major axis}                                                       &      &                                      \multicolumn{4}{c}{minor axis}                                                                           \\
\hline                                                                                                                                                                                                                                                                                                          
   $\log(r/r_{\rm eff})$            &    log(age)        &        [Fe/H]          &         [E/Fe]          &        [Z/H]      &      &        log(age)         &       [Fe/H]          &           [E/Fe]           &         [Z/H]     \\
\hline                                                                                                                                                                                                                                                                                                          
 $-$0.12                              &    1.175(0.148)                &     -0.50(0.18)                                  &       0.53(0.10)              &      0.00(0.13)                             &      &         1.175(0.134)                &       -1.03(0.24)                               &            0.80(0.32)            &        -0.28(0.23)                          \\
 $-$0.30                              &    1.025(0.033)                &     -0.37(0.07)                                  &       0.53(0.06)              &      0.13(0.04)                             &      &         1.150(0.106)                &       -0.34(0.15)                               &            0.36(0.09)            &         0.00(0.14)                          \\
 $-$0.46                              &    1.175(0.043)                &     -0.28(0.08)                                  &       0.30(0.09)              &      0.00(0.06)                             &      &         1.075(0.058)                &       -0.40(0.15)                               &            0.48(0.13)            &         0.05(0.07)                          \\
 $-$0.59                              &    1.175(0.030)                &     -0.41(0.07)                                  &       0.38(0.05)              &     -0.05(0.04)                             &      &         1.175(0.029)                &       -0.23(0.09)                               &            0.38(0.06)            &         0.13(0.05)                          \\
 $-$0.74                              &    1.175(0.039)                &     -0.27(0.05)                                  &       0.21(0.05)              &     -0.08(0.05)                             &      &         1.175(0.040)                &       -0.40(0.09)                               &            0.34(0.06)            &        -0.08(0.04)                          \\
 $-$0.92                              &    1.175(0.000)                &     -0.38(0.06)                                  &       0.40(0.05)              &      0.00(0.03)                             &      &         1.125(0.105)                &       -0.13(0.10)                               &            0.38(0.04)            &         0.23(0.09)                          \\
 $-$1.13                              &    1.025(0.060)                &     -0.15(0.07)                                  &       0.24(0.04)              &      0.08(0.07)                             &      &         1.125(0.040)                &       -0.13(0.07)                               &            0.32(0.04)            &         0.18(0.04)                          \\
 $-$1.38                              &    0.975(0.125)                &      0.00(0.09)                                  &       0.32(0.04)              &      0.30(0.07)                             &      &         1.050(0.064)                &        0.01(0.06)                               &            0.34(0.05)            &         0.33(0.05)                          \\
 $-$1.73                              &    0.900(0.133)                &      0.20(0.10)                                  &       0.21(0.04)              &      0.40(0.08)                             &      &         1.000(0.124)                &       -0.02(0.09)                               &            0.34(0.05)            &         0.30(0.07)                          \\
 Centre                               &    0.975(0.158)                &      0.15(0.06)                                  &       0.24(0.04)              &      0.38(0.06)                             &      &         0.950(0.103)                &        0.12(0.06)                               &            0.32(0.04)            &         0.43(0.05)                          \\
 $-$1.73                              &    0.900(0.107)                &      0.10(0.10)                                  &       0.21(0.06)              &      0.30(0.07)                             &      &         1.050(0.032)                &        0.02(0.05)                               &            0.38(0.04)            &         0.38(0.03)                          \\
 $-$1.38                              &    1.100(0.021)                &      0.02(0.05)                                  &       0.27(0.04)              &      0.28(0.02)                             &      &         1.050(0.046)                &        0.00(0.06)                               &            0.40(0.04)            &         0.38(0.03)                          \\
 $-$1.13                              &    1.075(0.011)                &     -0.13(0.06)                                  &       0.27(0.05)              &      0.13(0.05)                             &      &         1.075(0.047)                &        0.02(0.06)                               &            0.30(0.05)            &         0.30(0.03)                          \\
 $-$0.92                              &    1.175(0.000)                &     -0.27(0.06)                                  &       0.34(0.05)              &      0.05(0.04)                             &      &         1.175(0.019)                &       -0.21(0.05)                               &            0.30(0.04)            &         0.08(0.04)                          \\
 $-$0.74                              &    1.175(0.020)                &     -0.18(0.08)                                  &       0.24(0.05)              &      0.05(0.05)                             &      &         1.175(0.021)                &       -0.28(0.06)                               &            0.32(0.05)            &         0.03(0.02)                          \\
 $-$0.59                              &    1.150(0.049)                &     -0.31(0.07)                                  &       0.36(0.05)              &      0.03(0.04)                             &      &         1.175(0.021)                &       -0.22(0.08)                               &            0.34(0.06)            &         0.10(0.03)                          \\
 $-$0.46                              &    1.175(0.098)                &     -0.50(0.12)                                  &       0.34(0.08)              &     -0.18(0.07)                             &      &         1.175(0.033)                &       -0.31(0.13)                               &            0.36(0.11)            &         0.03(0.06)                          \\
 $-$0.30                              &    1.175(0.129)                &     -0.48(0.10)                                  &       0.38(0.10)              &     -0.13(0.09)                             &      &         1.175(0.145)                &       -0.33(0.10)                               &            0.40(0.08)            &         0.05(0.10)                          \\
 $-$0.12                              &    1.075(0.071)                &     -0.50(0.13)                                  &       0.53(0.09)              &      0.00(0.07)                             &      &         1.175(0.133)                &       -1.03(0.21)                               &            0.80(0.28)            &        -0.28(0.29)                          \\
\hline                                          
                                                
\hline                                          
\end{tabular}                                  
\label{data734}                                
\end{centering}
\end{table*}

\begin{table*}
\begin{centering}
\caption{Measured stellar population parameters for SDSS0758. Results are shown for both major and minor axes. Errors are given in brackets. Log(age) is expressed in Gyr.}
\begin{tabular}{cccccccccc}
\hline                                                                                                                                                                                                                                                                                                                                  
                                  &                                                       \multicolumn{4}{c}{major axis}                                                       &      &                                      \multicolumn{4}{c}{minor axis}                                                                           \\
\hline                                                                                                                                                                                                                                                                                                                                  
   $\log(r/r_{\rm eff})$          &        log(age)        &        [Fe/H]          &         [E/Fe]          &        [Z/H]      &      &        log(age)         &      [Fe/H]          &           [E/Fe]           &         [Z/H]     \\
\hline                                                                                                                                                                                                                                                                                                                                  
$-$0.08                             &           1.175(0.080)             &                 -0.34(0.20)                      &          0.12(0.18)           &                -0.23(0.09)                  &      &        1.100(0.167)                 &               $-$1.43(0.19)                       &          0.80(0.12)              &             -0.68(0.16)                     \\
$-$0.22                             &           1.175(0.092)             &                 -0.47(0.12)                      &          0.18(0.08)           &                -0.30(0.07)                  &      &        1.050(0.068)                 &               $-$0.40(0.08)                       &          0.21(0.08)              &             -0.20(0.04)                     \\
$-$0.37                             &           0.900(0.134)             &                 -0.36(0.16)                      &          0.09(0.12)           &                -0.28(0.08)                  &      &        1.175(0.023)                 &               $-$0.45(0.10)                       &          0.24(0.10)              &             -0.23(0.05)                     \\
$-$0.43                             &           1.100(0.041)             &                 -0.56(0.10)                      &          0.12(0.10)           &                -0.45(0.09)                  &      &        1.125(0.034)                 &               $-$0.35(0.10)                       &          0.27(0.10)              &             -0.10(0.05)                     \\
$-$0.51                             &           1.050(0.088)             &                 -0.28(0.10)                      &          0.24(0.09)           &                -0.05(0.05)                  &      &        1.175(0.040)                 &               $-$0.49(0.11)                       &          0.15(0.11)              &             -0.35(0.06)                     \\
$-$0.59                             &           1.175(0.016)             &                 -0.35(0.07)                      &          0.27(0.07)           &                -0.10(0.03)                  &      &        1.025(0.027)                 &               $-$0.43(0.07)                       &          0.27(0.08)              &             -0.18(0.05)                     \\
$-$0.70                             &           1.050(0.057)             &                 -0.30(0.07)                      &          0.24(0.06)           &                -0.08(0.05)                  &      &        1.050(0.040)                 &               $-$0.40(0.07)                       &          0.24(0.06)              &             -0.18(0.04)                     \\
$-$0.82                             &           1.175(0.021)             &                 -0.19(0.05)                      &          0.12(0.05)           &                -0.08(0.04)                  &      &        1.175(0.005)                 &               $-$0.30(0.07)                       &          0.27(0.08)              &             -0.05(0.06)                     \\
$-$0.93                             &           1.050(0.023)             &                 -0.21(0.06)                      &          0.30(0.06)           &                 0.08(0.04)                  &      &        1.125(0.041)                 &               $-$0.33(0.06)                       &          0.27(0.06)              &             -0.08(0.05)                     \\
$-$1.04                             &           1.125(0.008)             &                 -0.20(0.05)                      &          0.32(0.05)           &                 0.10(0.04)                  &      &        1.000(0.099)                 &               $-$0.15(0.08)                       &          0.34(0.05)              &              0.18(0.05)                     \\
$-$1.15                             &           1.075(0.019)             &                 -0.05(0.05)                      &          0.27(0.04)           &                 0.20(0.03)                  &      &        1.075(0.020)                 &               $-$0.24(0.06)                       &          0.36(0.05)              &              0.10(0.05)                     \\
$-$1.27                             &           1.050(0.016)             &                 -0.10(0.05)                      &          0.24(0.04)           &                 0.13(0.04)                  &      &        1.125(0.018)                 &               $-$0.25(0.05)                       &          0.34(0.05)              &              0.08(0.04)                     \\
$-$1.40                             &           1.175(0.000)             &                 -0.17(0.04)                      &          0.34(0.03)           &                 0.15(0.03)                  &      &        1.075(0.020)                 &               $-$0.10(0.05)                       &          0.34(0.03)              &              0.23(0.03)                     \\
$-$1.55                             &           1.125(0.011)             &                 -0.07(0.04)                      &          0.34(0.04)           &                 0.25(0.03)                  &      &        1.125(0.041)                 &               $-$0.16(0.07)                       &          0.36(0.05)              &              0.18(0.04)                     \\
$-$1.72                             &           1.050(0.043)             &                 -0.08(0.05)                      &          0.38(0.03)           &                 0.28(0.03)                  &      &        1.125(0.037)                 &               $-$0.18(0.05)                       &          0.32(0.04)              &              0.13(0.04)                     \\
$-$2.03                             &           1.125(0.000)             &                 -0.15(0.05)                      &          0.42(0.05)           &                 0.25(0.03)                  &      &        1.075(0.040)                 &               $-$0.15(0.05)                       &          0.34(0.04)              &              0.18(0.04)                     \\
Centre                              &           1.125(0.004)             &                 -0.18(0.05)                      &          0.40(0.04)           &                 0.20(0.03)                  &      &        1.075(0.039)                 &               $-$0.15(0.05)                       &          0.34(0.03)              &              0.18(0.04)                     \\
$-$2.03                             &           1.175(0.004)             &                 -0.21(0.04)                      &          0.38(0.04)           &                 0.15(0.03)                  &      &        1.125(0.014)                 &               $-$0.14(0.04)                       &          0.36(0.04)              &              0.20(0.04)                     \\
$-$1.72                             &           1.050(0.044)             &                 -0.16(0.06)                      &          0.38(0.05)           &                 0.20(0.04)                  &      &        1.050(0.116)                 &               $-$0.16(0.08)                       &          0.36(0.04)              &              0.18(0.06)                     \\
$-$1.55                             &           1.050(0.036)             &                 -0.06(0.06)                      &          0.38(0.04)           &                 0.30(0.04)                  &      &        1.125(0.037)                 &               $-$0.15(0.06)                       &          0.34(0.04)              &              0.18(0.03)                     \\
$-$1.40                             &           1.175(0.000)             &                 -0.03(0.04)                      &          0.27(0.04)           &                 0.23(0.02)                  &      &        1.075(0.048)                 &               $-$0.13(0.05)                       &          0.32(0.05)              &              0.18(0.04)                     \\
$-$1.27                             &           1.125(0.038)             &                 -0.25(0.05)                      &          0.34(0.04)           &                 0.08(0.04)                  &      &        1.050(0.044)                 &               $-$0.11(0.06)                       &          0.30(0.05)              &              0.18(0.04)                     \\
$-$1.15                             &           1.075(0.004)             &                 -0.10(0.04)                      &          0.32(0.04)           &                 0.20(0.03)                  &      &        1.175(0.007)                 &               $-$0.23(0.05)                       &          0.27(0.05)              &              0.03(0.04)                     \\
$-$1.04                             &           1.025(0.078)             &                 -0.13(0.07)                      &          0.27(0.05)           &                 0.13(0.06)                  &      &        0.975(0.082)                 &               $-$0.11(0.06)                       &          0.30(0.05)              &              0.18(0.05)                     \\
$-$0.93                             &           1.075(0.026)             &                 -0.10(0.07)                      &          0.27(0.06)           &                 0.15(0.05)                  &      &        1.150(0.014)                 &               $-$0.23(0.05)                       &          0.27(0.06)              &              0.03(0.03)                     \\
$-$0.82                             &           1.075(0.049)             &                 -0.31(0.08)                      &          0.30(0.07)           &                -0.03(0.04)                  &      &        1.125(0.024)                 &               $-$0.32(0.06)                       &          0.18(0.06)              &             -0.15(0.04)                     \\
$-$0.67                             &           1.000(0.090)             &                 -0.38(0.07)                      &          0.30(0.04)           &                -0.10(0.06)                  &      &        1.125(0.018)                 &               $-$0.43(0.05)                       &          0.27(0.05)              &             -0.18(0.04)                     \\
$-$0.55                             &           1.025(0.071)             &                 -0.20(0.10)                      &          0.27(0.07)           &                 0.05(0.06)                  &      &        1.000(0.150)                 &               $-$0.38(0.14)                       &          0.27(0.08)              &             -0.13(0.09)                     \\
$-$0.47                             &           1.175(0.100)             &                 -0.22(0.14)                      &          0.15(0.10)           &                -0.08(0.11)                  &      &        1.175(0.057)                 &               $-$0.35(0.16)                       &          0.18(0.13)              &             -0.18(0.07)                     \\
$-$0.39                             &           1.175(0.021)             &                 -0.56(0.16)                      &          0.24(0.10)           &                -0.33(0.06)                  &      &        1.075(0.041)                 &               $-$0.30(0.10)                       &          0.21(0.10)              &             -0.10(0.08)                     \\
$-$0.32                             &           1.075(0.143)             &                 -0.24(0.16)                      &          0.15(0.12)           &                -0.10(0.10)                  &      &        1.075(0.119)                 &               $-$0.24(0.16)                       &          0.12(0.15)              &             -0.13(0.09)                     \\
$-$0.21                             &           1.175(0.141)             &                 -0.45(0.19)                      &          0.21(0.15)           &                -0.25(0.14)                  &      &        1.175(0.080)                 &               $-$0.30(0.17)                       &          0.21(0.15)              &             -0.10(0.09)                     \\
$-$0.08                             &           0.275(0.143)             &                 -0.73(0.24)                      &          0.80(0.11)           &                 0.03(0.22)                  &      &        0.900(0.292)                 &               $-$0.37(0.24)                       &          0.15(0.18)              &             -0.23(0.19)                     \\
\hline                                               
\end{tabular}                                        
\label{data758}
\end{centering}
\end{table*}

\label{lastpage}

\end{document}